\documentclass[11pt]{article}
\usepackage[numbers]{natbib}
\usepackage{setspace}
\usepackage{float}
\usepackage{graphicx}
\usepackage{subfigure}

\usepackage{tkz-euclide}
\usepackage{tikz}

\usetikzlibrary{shapes,arrows,matrix}
\usepackage[linesnumbered,ruled]{algorithm2e}
\usepackage{multirow}
\usepackage{booktabs}

\usepackage{url}

\usepackage{footmisc}
\DefineFNsymbols{mySymbols}{{\ensuremath\dagger}{\ensuremath\ddagger}\S\P *{**}{\ensuremath{\dagger\dagger}}{\ensuremath{\ddagger\ddagger}}}
\setfnsymbol{mySymbols}
\usepackage{amssymb,amsmath,amsthm, amsfonts, mathtools}
\theoremstyle{definition} 
\newtheorem{problem}{Problem}

\begin{document}
\title{A 2D based Partition Strategy for Solving Ranking under Team Context (RTP)}

\author{
Xiaolu Lu\\
School of Software, Nanjing University\\
Jiangsu, China\\
mf1232050@software.nju.edu.cn
\and Dongxu Li\\
School of Software, Nanjing University\\
Jiangsu, China\\
       mf1332027@software.nju.edu.cn
\and  Xiang Li\thanks{Xiang Li is the corresponding author}\\
School of Software, Nanjing University\\
Jiangsu, China\\
       lx@software.nju.edu.cn
\and Ling Feng\\
Dept. of CS$\&$T, Tsinghua University\\
Beijing, China\\
       fengling@tsinghua.edu.cn\\
}
\maketitle
~\\
\begin{abstract}
In this paper, we propose a 2D based partition method for solving the problem of \textit{Ranking under Team Context(RTC)} on datasets without a priori. We first map the data into 2D space using its minimum and maximum value among all dimensions. Then we construct window queries with consideration of current team context. Besides, during the query mapping procedure, we can pre-prune some tuples which are not top ranked ones. This pre-classified step will defer processing those tuples and can save cost while providing solutions for the problem. Experiments show that our algorithm performs well especially on large datasets with correctness.
\end{abstract}
~\\

{\bf Keywords}: Team Context, Partition-based Index, Multidimensional Data
~\\

{\bf Topics} Database Application
~\\

\section{Introduction}
With the rapid development of context-aware computing techniques, context-aware applications are designed especially to enrich user experiences. They are modelled to generate results according to current situations\cite{DBLP:dblp_conf/comsnets/AmoghPAHF14}. However, most of the applications consider the context for users' individual preferences while researches are rarely conducted on group or team context\cite{DBLP:dblp_conf/icdim/LiF12a,DBLP:dblp_journals/eswa/PerezCH11,DBLP:dblp_conf/er/StefanidisSNK12}.

Team context are common in real world. Improving the competency is what the team leader really concerns about. Competency can be evaluated from various aspects. For example, if we set a team as a criterion, then competency of other teams can simply be measured using the distance between itself and the criterion. If a team context can become more nearer to a higher ranked one, it can be considered an improvement of competency. For example, assume the criterion team $T$ is the last one who has the ticket to enter into finals. Distance between a lower ranked team $A$ and $T$ is $d$. If distance between $A$ and $T$ is shortened, $A$ will increase the probability for entering into the finals. In this case, we might consider $A$ has an improvement in its competency. To this end, one common way is to exchange one member in the team with another. In \cite{lu2013object}, we define the problem of \textit{Ranking under Team Context} according to the situation. It aims at helping team leader to decide the swap-out team member and corresponding swap-in so that it can make current team becoming nearer to its target. For making decision about which object can be swap-in, we define a virtual object as an auxiliary. The virtual object take three aspects into consideration: current team context, the target and the swap-out object. It can be calculated using Equation. \ref{eq:vp calculation}.
\begin{equation}\label{eq:vp calculation}
v_i = \delta_i+r_i
\end{equation}
\noindent where $\delta_i$ is the difference between target and current team context, $r_i$ is the value of swap-out object on the \textit{i}th dimension.It has been proved in our previous work that objects with the highest rank are the nearest neighbours to an virtual object $V(v_1,v_2,..,v_d)$ 

With the sophisticated description of data as vectors, we will meet the so-called "curse of dimensionality". To solve this problem, researchers often exploit methods aiming at dimensionality reduction. Therefore, partition based methods are often used. Those strategies perform well on specific data\cite{DBLP:dblp_journals/tods/ZhangKOT05}, such as pyramid which has a high performance on uniformly distributed data. This technique chooses the center point of data space to divide \textit{d}-dimensional data space into pyramids with has size of \textit{2d}.\cite{berchtold1998pyramid}. Thus, non-uniformed datasets might cause performance degradation while applying the pyramid technique. Other partition based strategy such as \textit{i}Distance, which is only designed for effectively searching for \textit{k}NN. This \textit{state-of-art} technique will lose its advantage on non-clustered data.

If we simply want to make current team context nearer to its target, this means there is no specific requirement on which object to be swap-out. Denote $V_1$ as the corresponding virtual object of swap-out object $R_1$. Performing 1NN search based on query point $V_1$ will retrieve the swap-in object $P_1$. This result meets the requirement of minimizing distance between current and target team context and can be selected as a solution plan for the RTC problem. Obviously, different swap-out objects lead to different chosen swap-in objects. The one who shortens the distance in maximum is the top ranked tuple we prefer. So it is insufficient only using \textit{k}NN for providing the top ranked tuple. To be specific, \textit{k}NN search is just the first step in providing potential candidates in our problem. Moreover, what makes the RTC problem not so easy to be tackled, besides the \textit{curse of dimensionality}, is that we need to put all objects under team context, which varies from case to case. A more flexible plan for this problem needs to be developed. This plan needs to meet the potential requirement that can be applied to different team contexts flexibly. So basically, we want to design an approach which can (i)\textit{break the curse of dimensionality} to its best, (ii) minimize the cost in providing top ranked tuples for solving \textit{RTC} problem, and (iii) be adapted to different team context. Motivated by this, in this paper, we propose a plan for mapping the RTC query into a tunable query window based on a 2D partitioned strategy.  

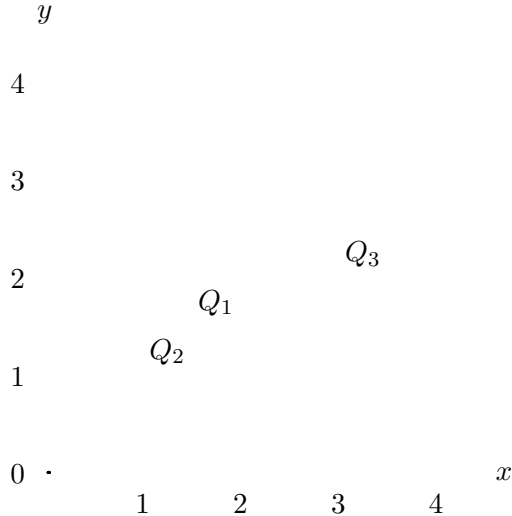
\begin{figure}\label{fig:tunable param of rec}
\center
\begin{tikzpicture}[scale=1.3]
\draw[dotted,step=.5cm] (0,0) grid (4.5,4.5);
\draw[dotted,step=.5cm] (0,0) grid (4.5,4.5);
\foreach \x in {1,...,4} { \node [anchor=north] at (\x,-0.1) {\x}; }
\foreach \y in {0,1,...,4} { \node [anchor=east] at (-0.1,\y) {\y}; }
    \draw [<->,thick] (0,4.5) node (yaxis) [above] {$y$}
        |- (4.5,0) node (xaxis) [right] {$x$};
    \draw [black] plot [only marks, mark size=1, mark=*] coordinates {(0.1,2.9)(1.1,1.9) (0.1,0.8)(2.1,1.8)(0.7,3)(2.1,2.9)(3.1,2.9)(1.7,0.9)(2.3,2.5)(2.4,0.6)(2.6,1.5)(1,2)};
    \draw[thick] (1.5,1.5) rectangle (2,2);
    \draw (1.75,1.75) node{$Q_1$};
     \draw[thick,dashed] (1,1) rectangle (1.5,1.5);
    \draw (1.25,1.25) node{$Q_2$};

     \draw[thick] (3,2) rectangle (3.5,2.5);
      \draw (3.25,2.25) node{$Q_3$};
    \end{tikzpicture}
\caption{Query Window under Different Team Context}
\end{figure}

Aiming at solving the RTC problem, we need to design an approach to effectively tackle the high dimensional data as the first step. Most of previous literatures have mapped the multidimensional data into single data space, but we adopt a strategy in another way. We mapped the data into 2D space using its maximum and minimum value. The rationale behind the intuition is based on the following two observations: (i)majority of the data can be identified using minimum and maximum value; (ii)if data is distributed nearer in the space established by their minimum and maximum value, they would like to have high probability becoming the nearest neighbour in a higher dimensional space. Take an example, if two students both scored best on math and worst on literature among all other subjects, we often think the two students are similar. In our mapping approach, they will be mapped in the same partition.
This means that we can reduce the dimensionality by avoiding the key collision which will incur in most of one-dimensional mapping schemes. Due to this, we propose a 2D partition based strategy in this paper as a basis for solving this problem. Data are mapped using both of the minimum and maximum value. The mapping key contains not only the value information but also partial dimension informations. Because RTC(Ranking under Team Context) problem needs to be solved for finding nearest neighbours, we map the RTC query to a window query in a 2D space. In designing the mapping technique, we incorporate the team context into consideration and provide a plan containing tunable parameters. Consider an example shown in Fig.\ref{fig:tunable param of rec}, the tunable context-aware parameters will first determine which partition to put the query window($Q_1$ or $Q_3$) and then where to put in the partition($Q_1$ or $Q_2$).
Therefore, parameters (i) serve as a filter in recognizing objects which cannot be the top ranked tuples. This filter saves the overhead cost in next \textit{NN} searching stage and (ii) decide the location of the query window. This step can improve the hit ratio in the process of candidates searching.

In this paper, our experiments on real data obtain satisfied results. At the stage of query window construct on, our algorithm will pre-recognize some objects whose nearest neighbours cannot make current team context closer than others'. Those results will be recognized before constructing and performing window queries.  Not until existing results cannot meet the users' requirement, they will not be processed. This deferred process might save partial overheads in producing the solution .

The rest of our paper is organized as follows: In Section \ref{Section:related work}, we review exiting techniques. In Section \ref{Section:preliminaries}, we introduce the background of our problem and solution. In Section \ref{section:index}, we present the method and in section \ref{section:algorithm} the detailed search algorithm. Section \ref{section:experiment} is the experiments and in section \ref{section:conclusion} we conclude our methods with potential future work.

\section{Related Works}\label{Section:related work}
As there are increasing requirements on providing high quality results for users, data objects are modelled in more complexed way with more and more attributes. Thus, multidimensional data has always required for efficient solutions in processing either range queries\cite{DBLP:conf/ideas/ChovanecK13} or \textit{k}NN queries\cite{Katayama:1997:SIS:253260.253347,Huang:2006:MKN:2165614.2165641}, centralized or decentralized\cite{DBLP:journals/geoinformatica/TsatsanifosSS13} in a distributed or cloud computing environment\cite{Papadopoulos:2011:ADI:2114498.2116236}.

No doubt that among all the query types, \textit{k}NN query has been widely applied in various real situations. Main techniques  fall into three categories: R-Tree based solution, hash based solution and partition based solution. 

Traditional \textit{R-tree based} index structures regarding to this problem have been discussed in bulk of literatures. The R-tree family such as TV-Tree\cite{Lin:1994:TIS:615204.615210}, X-Tree\cite{Berchtold:1996:XIS:645922.673502} and UB-Tree\cite{Bayer:1997:UBM:645965.674403} have been carefully studied and become the most popular solutions. TV Tree\cite{Lin:1994:TIS:615204.615210} used partial feature dimensions in indexing the data, UB-Tree\cite{Bayer:1997:UBM:645965.674403} took advantage of the z-order for coding the data on each dimension.  X-Tree\cite{Berchtold:1996:XIS:645922.673502} was more suitable for medium-sized multidimensional data\cite{PLTree}.

LSH(Locality Sensitive Hash)\cite{gionis1999similarity} proposed by \citeauthor{gionis1999similarity} was tailored to solving problem of similarity search in high dimensional space. Its hash function will maximize the probability of collision of two similar objects. To reduce the space cost, \citeauthor{Satuluri:2012:BLS:2140436.2140440} proposed a tunable LSH method called BayesLSH. Moreover, PLSH proposed in \cite{Sundaram:2013:SSS:2556549.2556574} extended the LSH to a parallel computing environment, which had a higher performance on large scale stream data.
%
Indexing high dimensional data based on \textit{partition strategy} can enhance the performance of NN searching and lots of previous work have discussed about this. Related partition strategy falls into two categories \textit{data-based partition} and \textit{space-based partition}. \textit{Data-based partition} strategy partitions data based on cluster technique while \textit{space-based} method partitions exploits statistical information of data.

The state-of-art exact \textit{k}NN search technique \textit{iDistance} proposed by \citeauthor{DBLP:dblp_journals/tods/JagadishOTYZ05} in \cite{DBLP:dblp_journals/tods/JagadishOTYZ05} mapped high dimensional data into 1D space and exploited both kinds of partition techniques for a searching process. It performed well on datasets either (i) showed explicit clustered feature or (ii) followed uniform distribution\cite{DBLP:dblp_conf/bncod/PillaiSBA13}. However, while data doesn't fit either of above-mentioned criteria, the approach showed less competitive.
Therefore, another tunable \textit{space-based} data partition method \textit{iMinMax($\theta$)} has been proposed in \cite{DBLP:journals/vldb/YuBOT04}.  Comparing to \textit{pyramid technique}, it can be adapted to various data distributions. Its main idea was to "push" data to dimension whose value is maximum or minimum among all the others. Partitions can be recognized through the maximum or minimum dimension and data can be "stretched" using tuning parameters defined in the method. This method can handle especially range queries with efficiency. Moreover, because its underlying structure is B-Tree, it can be integrated into existing systems easily. However, in real scenarios, data might have too many minimum values on the same dimension which will cause too many collisions while indexing and lead to performance bottleneck. Meanwhile, the "MaxEdge" will be same which might lead to an implicit distinguishing between partitions and deteriorated the efficiency of range query. Motivated by this, we want to explore new adaptive partition strategy performs well on NN searching.

Recent research conducted in \cite{PLTree} integrated features of tree like index structure and partition strategy. The PL-Tree proposed by \citeauthor{PLTree} recursively partitioned the data into hypercubes with labels and constructed tree dynamically. However, its main strength is in providing an efficient range query.

Besides concentrating on accuracy of resultset, researches in \cite{Pillai:2013:EHI:2526048.2526080,Cacheda:2011:IKN:2063576.2063939} customize \textit{k}NN query into different contexts for providing higher quality result. In this paper, we plan to design a method in solving the \textit{k}NN problem under team context defined in \cite{lu2013object}.

\section{Preliminaries}\label{Section:preliminaries}
\subsection{Background}
\textit{Team Context} defined in our previous work in \cite{lu2013object} was referred to a context constructed by multiple \textit{d}-dimensional data objects in data space $\mathcal{D}$. Assume team context $C$ was constructed by \textit{d}-dimensional objects with cardinality of \textit{m}. The problem of \textit{ranking under team context} is how to choose a swap-out object $R(r_1,r_2,...,r_d), R \in C(c_1,c_2,..,c_d)$ and a swap-in object $P(p_1,p_2,...,p_d), P\in \mathcal{D}$ to make $C$ becomes similar to its target team context $T(t_1,t_2,...,t_d)$.

If we measure the similarity using Euclidean distance, we can prove that the swap-in object $P$ chosen in $\mathcal{D}$ is the nearest neighbour of a virtual object $V(v_1,v_2,...,v_3)$ where $v_i = (t_i-c_i+r_i)\lambda_i$. $\lambda$ is an exchange parameter serves for measuring the contribution of swap-in objects on \textit{i}th dimension under current team context.
\subsection{Problem Formulation}.
In \cite{lu2013object}, we proved the \textit{ranking under team context} problem can be solved for finding the nearest neighbour of a virtual object. We also conduct several experiments using heuristic methods and observed that the problem can be worked out using a NN query.

Define the distance between a \textit{m} sized team context $C$ and $T$ as $\mathrm{Dist}_{C,T}$. The team context $C$ after exchange one object is denoted as $C'$. Formally, our problem can be defined as:
\begin{problem}
Determine a set of swap-in objects $\mathbb{P}\{P_1,P_2,...,P_k\}$ correspond to a set of swap-out objects $\mathbb{R}\{R_1,r_2,...,R_k\},(k \leq m)$ of $C$, which makes $\mathrm{Min}(\mathrm{Dist}_{C',T}-\mathrm{Dist}_{C,T})$ after making an exchange with $<R_i,P_i>(R_i\in\mathbb{R},P_i\in\mathbb{P})$.
\end{problem}
However, unlike traditional \textit{k}-NN query which only needs to determine the \textit{k} nearest neighbours of query points, our problem needs to find \textit{k} top ranked tuples to make current team approaching its target.

\section{Indexing under Team Context}\label{section:index}
\begin{figure}
  \centering
    \includegraphics[scale=0.5]{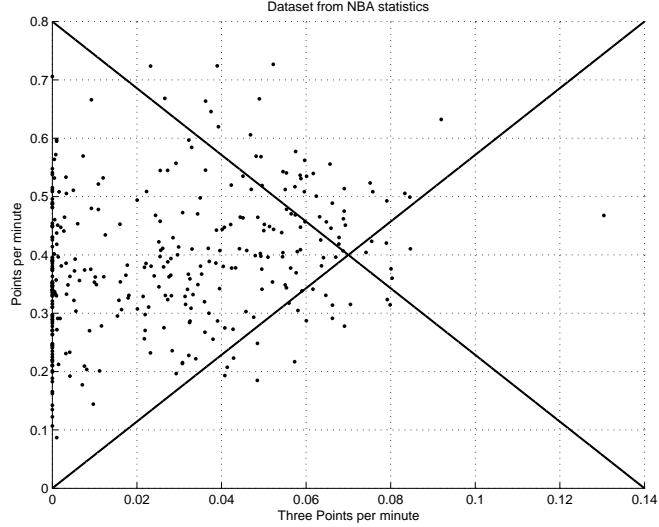}
  \caption{Distribution of NBA Dataset}
  \label{fig:Data distributions}
\end{figure}

In real cases the data distribution might be a skewed one as shown in \ref{fig:Data distributions}. This data shows less clustered feature and will be hard to "stretched" using \textit{iMinMax($\theta$)}\cite{DBLP:journals/vldb/YuBOT04}. Therefore, we consider to design a index technique which can distinguish data in multidimensional space and solve the problem effectively.
No doubt that NN query provides results which have the most similarity to query point. So intuitively, if both minimum value and maximum value of an object are close to a query point, it would have the most possibility to become a nearest neighbour correspondingly. Therefore, the strategy proposed in this paper uses both the minimum and maximum value of a multidimensional data object as a index key. Moreover, we hope that we can make the most of the strategy for distributing the data into different partitions. Since data are partitioned in a 2D space, we can easily generalized the information by location of data in this transformed space. Based on this observation, we plan to use \textit{quad tree} as the implementation structure.
\subsection{Mapping to a 2D Space}
In our paper, we consider a dataset $\mathcal{D}$ with \textit{d} dimensions. Each data object $R(r_1,r_2,...,r_d) \in \mathcal{D}$ has a maximum value $r_{max}= \mathrm{Max}_{i=1}^dr_i$ on dimension \textit{i}($i\leq d$) and minimum value $r_{min}=\mathrm{Min}_{j=1}^dr_j$ on dimension \textit{j}($j \leq d$). 
As introduced before, we use both minimum and maximum value of a data object as a index key. So key of a \textit{d-}dimensional data $R(r_1,r_2,...,r_d)$ is calculated as:
 \begin{equation}\label{eq:coordinate}
    \left\{
     \begin{array}{lr}
       x = i \times c + r_i, & r_i = \mathrm{Min}_{n=1}^dr_n\\
       y = j \times c + r_j, & r_j = \mathrm{Max}_{n=1}^dr_n
     \end{array}
   \right.
 \end{equation}
Notice that $c$ in Equation \ref{eq:coordinate} is a tunable parameter which can map multidimensional data into different partitions. $r_i$ refers to the minimum value of $R$ among all dimensions and $r_j$ refers to the maximum value. For example, consider a data point $A(0.1,0.25,0.7,0.9)$ in 4-dimensional space. If we set $c=0$,the key of $A$ is $(0.1,0.9)$, while $(1.1,4.9)$ if $c=1$. Notice that in this case, if we set $c=1$, we could obtain not only the minimum and maximum value among all dimensions of $A$, but also the information of the dimensions correspondingly. 

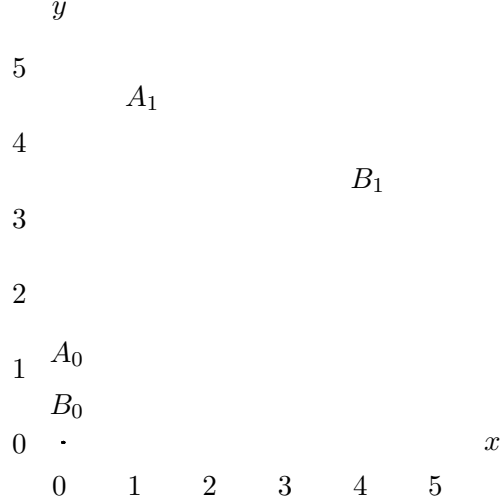
\begin{figure}
\center
\begin{tikzpicture}
\draw[dotted,step=.5cm] (0,0) grid (5,5);
\draw[dotted,step=.5cm] (0,0) grid (5,5);
    \draw [<->,thick] (0,5.5) node (yaxis) [above] {$y$}
        |- (5.5,0) node (xaxis) [right] {$x$};
\foreach \x in {0,1,...,5} {     
\draw [thick] (\x,0) -- (\x,-0.2);
}
\foreach \y in {0,1,...,5} {   
\draw [thick] (0,\y) -- (-0.2,\y);
}
\foreach \x in {0,1,...,5} { \node [anchor=north] at (\x,-0.3) {\x}; }
\foreach \y in {0,1,...,5} { \node [anchor=east] at (-0.3,\y) {\y}; }
\foreach \x in {.5,1.5,...,4.5} {
\draw [thin] (\x,0) -- (\x,-0.1);
}
\foreach \y in {.5,1.5,...,5.5} {
\draw [thin] (0,\y) -- (-0.1,\y);
}
\draw [black] plot [only marks, mark size=1.5, mark=*] coordinates {(0.1,0.9) (1.1,4.9) (0.1,0.8)(4.1,3.8)};
\draw (0.1,0.9) node[above]{$A_0$};
\draw (1.1,4.9) node[below]{$A_1$};
\draw (0.1,0.8) node[below]{$B_0$};
\draw (4.1,3.8) node[below]{$B_1$};
\end{tikzpicture}
\caption{Mapping with different $c$}\label{fig:map different c}
\end{figure}

Consider another data object $B(0.3,0.5,0.8,0.1)$ in 4-dimensional space. If we use $c=0$ to obtain the mapping key $A_0,B_0$ and $c=1$ to obtain $A_1,B_1$ respectively. Obviously, we can generalized from Fig.\ref{fig:map different c} that different value of $c$ will lead to different distributions of $A$ and $B$ in mapping space.

Parameter $c$ can be tuned to match the specific requirements. Through changing $c$, we can distribute the data with most similarity nearby.

Notice that the mapping process actually distribute the data into different locations in the 2D space. If we use $c=1$, we can obtain the dimension information and corresponding value easily from the mapping key. Moreover, since we do not using statistic information, so the mapping method can be adapted to different data distributions. Also, since a 2D indexing key was used, it could reduce the collisions to its best effort.

\subsection{Mapping the \textit{k}NN Query}
Since efficient method of \textit{i}Distance performs well on dataset either has a clustered feature or follows the uniform distribution, we cannot simply port the technique to our case. As discussed in many researches, \textit{k}NN query will often be transformed into range queries and adopt a \textit{filter-and-refine step} for seeking an approximate resultset.

Our problem is to find NN of $V(v_1,v_2,...,v_d)$ to serve as the swap-in object as a substitution of swap-out object $R$ under team context $C$. Generally, assume the search radius is $\Delta$, the corresponding range query is:
 \begin{equation}\label{eq:range query}
    \left\{
     \begin{array}{lr}
       U_i = v_i-\Delta\\
       L_i = v_i+\Delta
     \end{array}
   \right.
 \end{equation}
 \noindent where $U_i$ is the upper bound value on \textit{i}th dimension and $L_i$ is the lower bound value. However, to determine the mapped range query under current team context will be a little bit different, which will be introduced later.

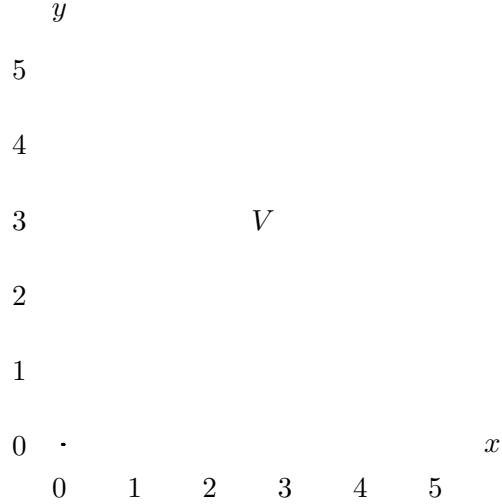
\begin{figure}
\center
\begin{tikzpicture}
\draw[dotted,step=.5cm] (0,0) grid (5,5);
\draw[dotted,step=.5cm] (0,0) grid (5,5);
    \draw [<->,thick] (0,5.5) node (yaxis) [above] {$y$}
        |- (5.5,0) node (xaxis) [right] {$x$};
\foreach \x in {0,1,...,5} {     
\draw  (\x,0) -- (\x,-0.2);
}
\foreach \y in {0,1,...,5} {   
\draw  (0,\y) -- (-0.2,\y);
}
\foreach \x in {0,1,...,5} { \node [anchor=north] at (\x,-0.3) {\x}; }
\foreach \y in {0,1,...,5} { \node [anchor=east] at (-0.3,\y) {\y}; }
\foreach \x in {.5,1.5,...,4.5} {
\draw [thin] (\x,0) -- (\x,-0.1);
}
\foreach \y in {.5,1.5,...,5.5} {
\draw [thin] (0,\y) -- (-0.1,\y);
}
\draw [black] plot [only marks, mark size=1, mark=*] coordinates {(3.1,2.9)(1.1,4.9) (0.1,0.8)(4.1,3.8)(3.7,3.9)(2.1,2.9)(4.1,2.9)(2.7,3.9)(2.3,3.7)(2.4,3.6)(2.6,3.6)(3,3)};
\draw[thick] (3,3) circle (0.8);
\draw[thick] (2.2,2.2) rectangle (3.8,3.8);
\draw (3,3) node[left]{$V$};
\end{tikzpicture}
\caption{Range query and NN query}\label{fig:rangeq and nnq}
\end{figure}

As shown that the mapped window query is a square which contains more candidate data objects than real needed as depicted in Fig.\ref{fig:rangeq and nnq}. It resulted from the additional enlarged area since we transform the original circle region to a square one. So a refine step is needed for removing the false hit.

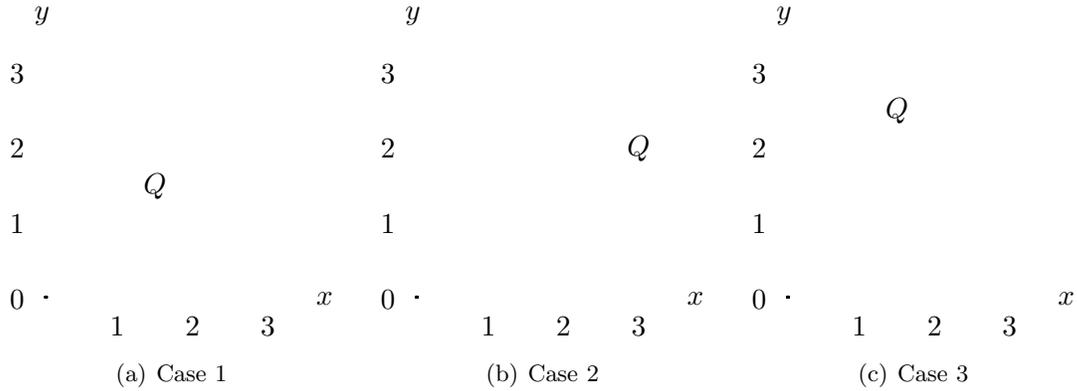
\begin{figure}
\center
 \subfigure[Case 1]{ 
    \label{fig:subfig:case1} 
   \begin{tikzpicture}
\draw[dotted,step=.5cm] (0,0) grid (3.5,3.5);
\draw[dotted,step=.5cm] (0,0) grid (3.5,3.5);
\foreach \x in {1,...,3} { \node [anchor=north] at (\x,-0.1) {\x}; }
\foreach \y in {0,1,...,3} { \node [anchor=east] at (-0.1,\y) {\y}; }
    \draw [<->,thick] (0,3.5) node (yaxis) [above] {$y$}
        |- (3.5,0) node (xaxis) [right] {$x$};
    \draw [black] plot [only marks, mark size=1, mark=*] coordinates {(0.1,2.9)(1.1,1.9) (0.1,0.8)(2.1,1.8)(0.7,3)(2.1,2.9)(3.1,2.9)(1.7,0.9)(2.3,2.5)(2.4,0.6)(2.6,1.5)(1,2)};
    \draw[thick] (1,1) rectangle (2,2);
    \draw (1.5,1.5) node{$Q$};
    \end{tikzpicture}
    } 
  \subfigure[Case 2]{ 
    \label{fig:subfig:case2} 
   \begin{tikzpicture}
\draw[dotted,step=.5cm] (0,0) grid (3.5,3.5);
\draw[dotted,step=.5cm] (0,0) grid (3.5,3.5);
\foreach \x in {1,...,3} { \node [anchor=north] at (\x,-0.1) {\x}; }
\foreach \y in {0,1,...,3} { \node [anchor=east] at (-0.1,\y) {\y}; }
    \draw [<->,thick] (0,3.5) node (yaxis) [above] {$y$}
        |- (3.5,0) node (xaxis) [right] {$x$};
    \draw [black] plot [only marks, mark size=1, mark=*] coordinates {(0.1,2.9)(1.1,1.9) (2.1,1.8)(0.7,3)(3.1,2.9)(1.7,0.9)(2.3,2.5)(2.6,2.8)(1,2)};
    \draw[thick] (2.5,1.5) rectangle (3.5,2.5);
    \draw (3,2) node{$Q$};
    \end{tikzpicture}
    }
   \subfigure[Case 3]{ 
    \label{fig:subfig:case3} 
   \begin{tikzpicture}
\draw[dotted,step=.5cm] (0,0) grid (3.5,3.5);
\draw[dotted,step=.5cm] (0,0) grid (3.5,3.5);
\foreach \x in {1,...,3} { \node [anchor=north] at (\x,-0.1) {\x}; }
\foreach \y in {0,1,...,3} { \node [anchor=east] at (-0.1,\y) {\y}; }
    \draw [<->,thick] (0,3.5) node (yaxis) [above] {$y$}
        |- (3.5,0) node (xaxis) [right] {$x$};
    \draw [black] plot [only marks, mark size=1, mark=*] coordinates {(0.1,2.9)(2.1,2.9) (2.1,1.8)(0.7,3)(3.1,2.9)(2.7,0.9)(2.3,2.5)(2.6,2.8)(1,1.5)(2.5,2)};
    \draw[thick] (1,2) rectangle (2,3);
    \draw (1.5,2.5) node{$Q$};
    \end{tikzpicture}
    }
\caption{Query Window Cases} \label{fig:query window}
\end{figure}

Mapping NN query contains two steps: \textit{transforming to range query} and \textit{mapping into 2D space}. Decide how to map the dimensions into sub queries in 2D space is the key step. Intuitively, we should map the query into $C_{d}^2$ query windows. However, not all the query windows are necessary. There might exist three cases as shown in Fig.\ref{fig:query window}.

\textbf{Case 1} As depicted in Fig.\ref{fig:subfig:case1}, the window query can cover set of points. Those points covered by window query are potential candidates. They can be further refined for achieving the needed result.

\textbf{Case 2} In this case shown in Fig.\ref{fig:subfig:case2}, the query window contains no possible results. Recall that we already assume that values among all dimensions of our data are in the range [0,1]. So query window $Q$ actually covers 3 partitions. Clearly, $Q$ requires range from [2.5,3.5] on \textit{x} dimension and [1.5,2.5] on \textit{y}. This means it queries a data has minimum value on 2nd dimension with ranging from 0.5 to 1 or data with minimum value on 3th dimension whose value falls in [0,0.5]. Also, the maximum value of such data should fulfil a value requirement of  [0.5,1] on 1st dimension or [0,0.5] on the 2nd dimension. Although Q contains nothing in this case, this mapped window query cannot be pre-pruned. In this case, we would need to enlarge query window in the following steps.

\textbf{Case 3} Notice that this case illustrated in Fig.\ref{fig:subfig:case3} seems alike to previous one. However, query window Q in this case is ([1,2],[2,3]). This means that no data in our dataset would has minimum value on dimension 1 while has the maximum value on dimension 2. This case can be identified at the beginning of query mapping, thus we can perform a pre-prune step before starting the range query.

Denote set which contains information of all minimum dimensions as $\mathcal{M}$ and set which has information regarding to maximum dimensions as $\mathcal{H}$. So the mapped dimension set $X$ on \textit{x} coordinate and $Y$ on \textit{y} should meet the requirement as $X\subseteq\mathcal{M} \wedge Y\subseteq H$.

If search using the radius of $\Delta$, so the upper bound $U_i$ of query window on \textit{i}th dimension in the 2D space would be like:

 \begin{equation}\label{eq:query window upper}
    \left\{
     \begin{array}{lr}
       U_i = v_i+\Delta, \qquad v_i+\Delta\leq (i+1)\times c\\
       U_i = (i+1)\times c, \qquad v_i+\Delta> (i+1)\times c
     \end{array}
   \right.
 \end{equation}
\noindent the lower bound of the query can be calculated respectively.

\textbf{Mapping under Team Context}. Because we are dealing with the problem under a team context, so the dimensions chosen step and pre-prune step can be improved accordingly. If we choose \textit{m} dimensions as minimum dimensions to be mapped and \textit{n} as the maximum ones($m+n<d$). If we use $\mathcal{V}_{min,k}, \mathcal{V}_{max,k},\mathcal{R}_{min,k}, \mathcal{R}_{max,k}$ to denote the dimensions whose values are k minimum or maximum ones comparing to others of virtual object $V$ and $R$.  The mapped dimensions are determined as:
\begin{equation}\label{eq:setx and sety}
    \left\{
     \begin{array}{lr}
       X = \mathcal{V}_{min,m}\cap\mathcal{R}_{min,m}\\
       Y = \mathcal{V}_{max,n}\cap\mathcal{R}_{max,n}
     \end{array}
   \right.
 \end{equation}
\noindent There might exist such situation that $Y=\varnothing$. In such situation, we might downgrade the priority of $R$ to become a proper swap-out candidate who can make current team context approaching its target.
 
\section{Search Method}\label{section:algorithm}
In our implementation, we utilize quad-tree as our underlying index structure. However, since our problem mainly focus on determining the potential \textit{k} exchange pairs of current team context to make it nearer to its target, we will concentrate on this.
\subsection{Dimension Chosen}\label{subsection:dimension choose}
As described in previous section, we determine the dimensions which needed to be mapped using both swap-out object and virtual object. In Algorithm \ref{algo:choose dim}, we input both $V$ and $R$, sort them ascendantly and obtain their original dimension index after sort operation. \textit{m} and \textit{n} are dimensions needed in mapping process. Notice that we also utilize the available minimum and maximum dimension information $\mathcal{M}$ and $\mathcal{H}$ which can help us make a pre-prune on mapping the query.
\begin{algorithm}\label{algo:choose dim}
\SetAlgoNoLine
\caption{ChooseDimension}
\KwIn{$V,R$, set of minimum dimensions$\mathcal{M}$,set of maximum dimensions $\mathcal{H}$,\textit{m,n}}
\KwOut{$X,Y$}
$I_v \leftarrow$ Sort($V$)\;
$I_r \leftarrow$ Sort($R$)\;
$X \leftarrow I_v[1 \sim m] \cap I_r[1\sim m] \cap \mathcal{M}$\;
$Y \leftarrow I_v[d-n \sim d] \cap I_r[d-n \sim d] \cap \mathcal{H}$\;
return $X,Y$\;
\end{algorithm}

\subsection{The Query Window}
Searching the \textit{k}NN of query point $V$ is transformed into a range query and mapped into the 2D space accordingly. However, since we assume the query point $V$ is not in the dataset, which means while transforming the NN query, the center of the window query cannot be determined in some case(negative value appears, for example). Therefore, there are two possible query plans \textit{query on the grid} or \textit{query by offset}.
\subsubsection{The Position of Query Window}
\textit{query on the grid} can be simply viewed as a special case of \textit{query by offset} while the offset value $\theta=0$. So generally we just use $\theta$ as a tuning parameter which could be customized case by case. The parameter is used for fix the position of query window, which serves for speeding up searching procedure.

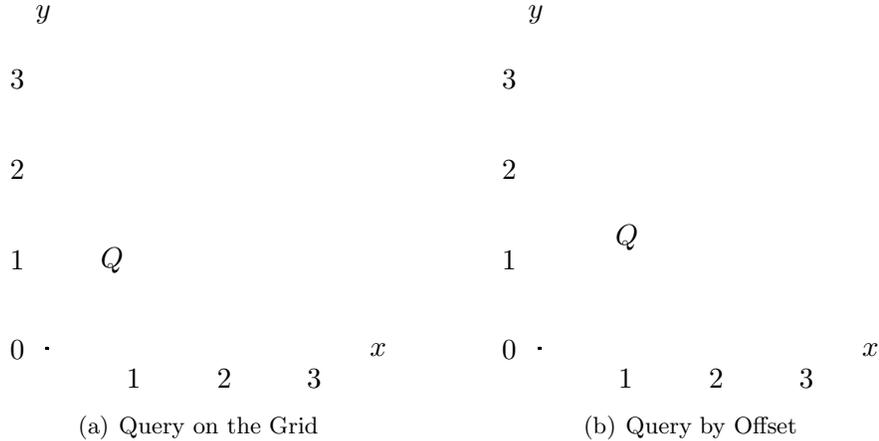
\begin{figure}
\center
 \subfigure[Query on the Grid]{ 
    \label{fig:subfig:pcase1} 
   \begin{tikzpicture}[scale=1.2]
\draw[dotted,step=.5cm] (0,0) grid (3.5,3.5);
\draw[dotted,step=.5cm] (0,0) grid (3.5,3.5);
\foreach \x in {1,...,3} { \node [anchor=north] at (\x,-0.1) {\x}; }
\foreach \y in {0,1,...,3} { \node [anchor=east] at (-0.1,\y) {\y}; }
    \draw [<->,thick] (0,3.5) node (yaxis) [above] {$y$}
        |- (3.5,0) node (xaxis) [right] {$x$};
    \draw [black] plot [only marks, mark size=1, mark=*] coordinates {(0.1,2.9)(1.1,1.9) (0.1,0.8)(2.1,1.8)(0.7,3)(2.1,2.9)(3.1,2.9)(1.7,0.9)(2.3,2.5)(2.4,0.6)(2.6,1.5)(1,2)(1.5,1.6)};
    \draw[thick] (1,1) rectangle (1.5,1.5);
    \draw (1,1) node[left]{$Q$};
    \end{tikzpicture}
    }
    \hspace{0.3in}
  \subfigure[Query by Offset]{ 
    \label{fig:subfig:pcase2} 
   \begin{tikzpicture}[scale=1.2]
\draw[dotted,step=.5cm] (0,0) grid (3.5,3.5);
\draw[dotted,step=.5cm] (0,0) grid (3.5,3.5);
\foreach \x in {1,...,3} { \node [anchor=north] at (\x,-0.1) {\x}; }
\foreach \y in {0,1,...,3} { \node [anchor=east] at (-0.1,\y) {\y}; }
    \draw [<->,thick] (0,3.5) node (yaxis) [above] {$y$}
        |- (3.5,0) node (xaxis) [right] {$x$};
    \draw [black] plot [only marks, mark size=1, mark=*] coordinates {(0.1,2.9)(1.1,1.9) (2.1,1.8)(0.7,3)(3.1,2.9)(1.7,0.9)(2.3,2.5)(2.6,2.8)(1,2)(1.5,1.6)};
    \draw[thick] (1.25,1.25) rectangle (1.75,1.75);
    \draw (1.25,1.25) node[left]{$Q$};
    \end{tikzpicture}
    }
\caption{Position of Query Window} \label{fig:position of query window}
\end{figure}

For example, if the query point has negative value as the minimum value, then we need to decide which way to perform the query in the 2D space. If we tackle the case using directly mapping, then the query will be performed from the partition edge, as shown in Fig.\ref{fig:subfig:pcase1}. So we need to iteratively enlarge the query window for finding the answer. However, if we set the offset parameter and use the way of querying by offset in Fig.\ref{fig:subfig:pcase2}, we can obtain the answer with less overhead.

Since the operation is performed under team context, this $\theta$ is context-related. It can be tuned vary from context to context.
\subsubsection{The Size of Query Window}
Since the search process transforms original \textit{k}NN search in to a window query on 2D space. Thus, the search radius is changing iteratively. Size of the query window are defined by initial search radius $\Delta$ and enlarging radius $\gamma$.

We first consider the case that the dimensions chosen in Section \ref{subsection:dimension choose} is $\varnothing$. Then the window size will be 0. This means the object cannot be swap-out to make current team becomes closer to its target. This can be used as a pre-prune step in \ref{algo:search candidate}, the object will be assigned a lower priority in the searching stage.

If we can get desired number of results, the enlarging step of query window will stop. Admittedly, this search operation may exceeds the partition border, so we might define a threshold. If the search exceeds the threshold when enlarging, then we consider this swap-out object is not a better under current context. Thus, the object will be in a lower priority for providing solutions same as the one whose $Y=\varnothing$.

\subsection{Search for Global Solution}
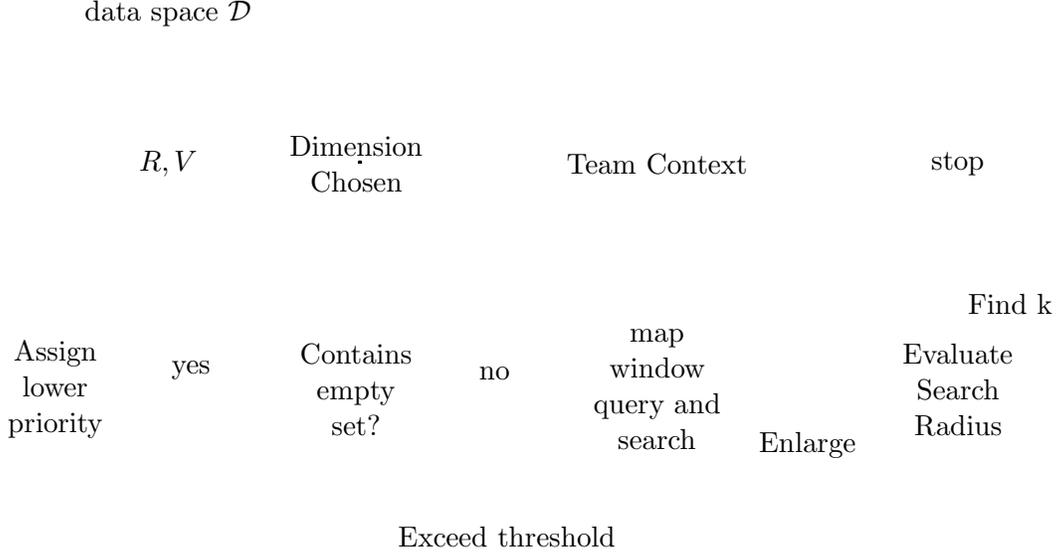
\begin{figure}
\center
\tikzstyle{decision} = [diamond, draw,
    text width=4.5em, text badly centered, node distance=3cm, inner sep=0pt]
\tikzstyle{block} = [rectangle, draw, node distance=2cm,
    text width=5em, text centered, rounded corners, minimum height=4em]
\tikzstyle{line} = [draw, -latex']
\tikzstyle{cloud} = [draw, ellipse, node distance=2.5cm,
    minimum height=2em]
    
\begin{tikzpicture}
    \node [block] (init) {Dimension Chosen};
    \node [cloud, left of=init] (expert) {$R,V$};
    \node [cloud, above of=expert,node distance=2cm] (data) {data space $\mathcal{D}$};
    \node [cloud, right of=init,node distance=4cm] (context) {Team Context};
    \node [decision, below of=init] (identify) {Contains empty set?};
    \node [block, left of=identify,node distance=4cm] (evaluate){Assign lower priority};

    \node [block, right of=identify, node distance=4cm] (map) {map window query and search};
    \node [block, right of=map,node distance=4cm] (decide) {Evaluate Search Radius};
    \node [block, above of=decide, node distance=3cm] (stop) {stop};
    \path [line,dashed] (expert) -- (init);
    \path [line] (init) -- (identify);
    \path [line,dashed] (data) -| (init);
    \path [line,dashed] (context)--(init);
    \path [line] (identify) --node [midway, above] {yes} (evaluate);
    \draw[line] (decide.south) to  [out=-165,in=-15] node [midway, below] {Exceed threshold} (evaluate.south);
    \draw[line] (decide) to  [out=-165,in=-15] node [midway, below] {Enlarge} (map);
        \draw[line] (map.east) to  [out=15,in=165] node [near start] {} (decide.west);
    \path[line] (decide) to  node [near start,right] {Find k} (stop);
    
      \path [line] (identify) --node [near start,above] {no} (map);

\end{tikzpicture}
\caption{Search under Team Context}\label{fig:search under team context}
\end{figure}

Assume that we choose $R$ as a swap-out object. The main steps in Algorithm\ref{algo:search candidate} can be simply illustrated in Fig.\ref{fig:search under team context}. First decide the dimensions for the mapping of query window, then perform either \textit{query by offset} or \textit{query on the grid}. This process repeats until we could find satisfied results.

\begin{algorithm}\label{algo:search candidate}
\SetAlgoNoLine
\caption{SearchCandidates}
\KwIn{$R,V,\mathcal{H},\mathcal{M}$,search radius $\Delta$, enlarge radius $\gamma$, offset value $\theta$, candidates number \textit{k}}
\KwOut{$\mathbb{P}$}
$X,Y \leftarrow \mathrm{ChooseDimension}(R,V)$\;
\If{$Y = \varnothing$}{ReduceSearchPriority(R)\;return\;}
\ForEach{dimension $i \in X$}
{\ForEach{dimension $j \in Y$}
{
$\mathrm{QueryCenter_x} \leftarrow i\times c+\theta$\;
$\mathrm{QueryCenter_y} \leftarrow j\times c+\theta$\;
$\mathrm{WindowQuery}_i \leftarrow$ TransformTo2D($\mathrm{QueryCenter_x,QueryCenter_y},\Delta$)\;
}
}
\ForEach{Key $K \in \mathrm{WindowQuery}$}
{
\lIf {refine($K,\Delta$)}{add \textit{K} to $\mathbb{P}$ }\;
}
\lIf{size$(\mathbb{P})<k$}{EnlargeQueryWindow($\gamma$)}\;
\If{EnlargeQueryWindow($\gamma$)$\geq \mathrm{threshold}$}{ReduceSearchPriority(R)\;return\;}
return $\mathbb{P}$\;
\end{algorithm}

WindowQuery described in line 10 will retrieves all data points covered and find the potential candidates. Notice that line 2 describes a pre-prune step as described previously. Perform the search method for every objects under team context until we could find \textit{k} answers. However, if all the objects has been processed and yet not find enough ones, we will going to check the objects with lower priority assigned before.


\section{Experiment}\label{section:experiment}
\subsection{Experiment Setup}
All the experiments were performed on machine with Intel Core(TM) i3 CPU and 4 GB RAM hosted on 32 bit Windows 7.

\textbf{Datasets}. We perform our experiments on real data obtained from \cite{website:basketball-reference} which consists total statistic data of NBA regular season 2011 $\sim$ 2012.  Real dataset contains 400 players with 24 attributes in total and and 30 teams described by 20 dimensions. Size of player dataset is 39.5KB and team dataset is 20KB.

Attributes which can discriminate between season-long successful and unsuccessful basketball teams according to researches on basketball in \cite{oliver2004basketball} are FG, 3P, 3PA, BLK, FT, STL, FTA, PTS, AST, DRB and TRB. We use this attribute set for our experiments as well.

\begin{figure}
  \centering
    \includegraphics[scale=0.5]{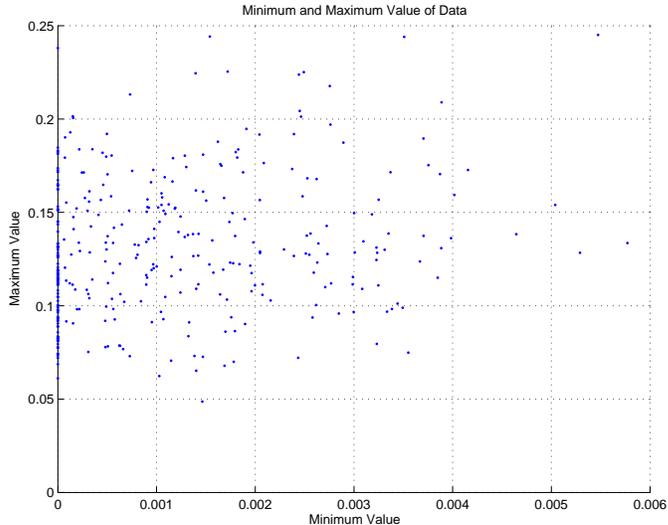}
  \caption{Overview of Min and Max Value}
  \label{fig:Overview of Each Dimension}
\end{figure}

We consider the statistics of player per minute since it could describe the feature of a player more properly. An overview of data description are listed as:

\begin{table}[htbp]
  \centering
  \caption{Overview of Data}
      \begin{tabular}{c|cccccc}
    \toprule
    Attribute & FG    & ThreeP & ThreePA & FT    & FTA   & PTS \\
    \midrule
    Minimum & 0.0107 & 0     & 0     & 0     & 0.0006 & 0.0292 \\
    Maximum & 0.0859 & 0.0197 & 0.0219 & 0.0235 & 0.0405 & 0.245 \\
    \hline
    Attribute & DRB   & TRB   & AST   & STL   & BLK   &  \\
    \hline
    Minimum & 0.0216 & 0.0194 & 0.0012 & 0.0008 & 0     &  \\
    Maximum & 0.1306 & 0.1293 & 0.0622 & 0.011 & 0.0257 &  \\
    \bottomrule
    \end{tabular}%
  \label{table:Overview of Data}%
\end{table}
An overview of minimum and maximum data amount on each dimension on real dataset is shown in Fig.\ref{fig:Overview of Each Dimension}. The synthetic data are generated according to the feature of real data set. It contains 1 million tuples with total size of 16M. 

\subsection{Efficiency on Finding \textit{k}NN}
We test our method on both real and synthetic dataset. Query point are calculated based on the definition of virtual objects. Details are listed in \ref{tab:query params}

\begin{figure}[H]
  \centering
    \includegraphics[scale=0.4]{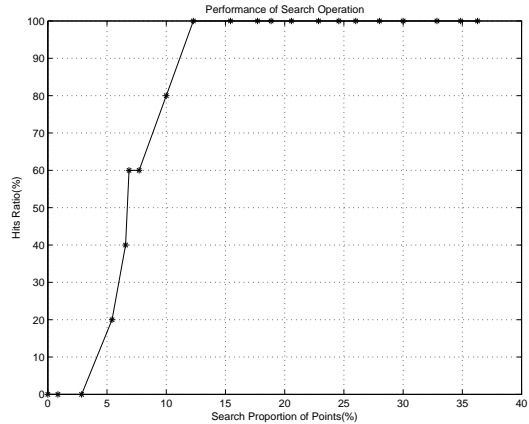}
  \caption{Search Proportions and Hits Ratio of 5NN on Real Data}
  \label{fig:5NN search}
\end{figure}

\begin{table}[htbp]
  \centering
  \caption{Experiment Parameters}
    \begin{tabular}{cccccc}
    \toprule
    parameter & $\gamma$ & k     & threshold & xoffset & yoffset \\
    \midrule
    value & 0.001 & 5     & 0.02  & 0.0013 & 0.1553 \\
    \bottomrule
    \end{tabular}%
  \label{tab:query params}%
\end{table}%
Figure \ref{fig:5NN search} shows NN search operation while \textit{k}=5. Hit ratio is the exact results contained in the query window to \textit{k}. Search proportion is the data queried in the query window to the cardinality of whole data space $\mathcal{D}$.
%

Notice that only small proportion of data needed to be queried for achieving desired results within the threshold. 
Notice that only small proportion of data needed to be queried for achieving desired results within the threshold. 
\begin{figure}[H]
  \centering
    \includegraphics[scale=0.4]{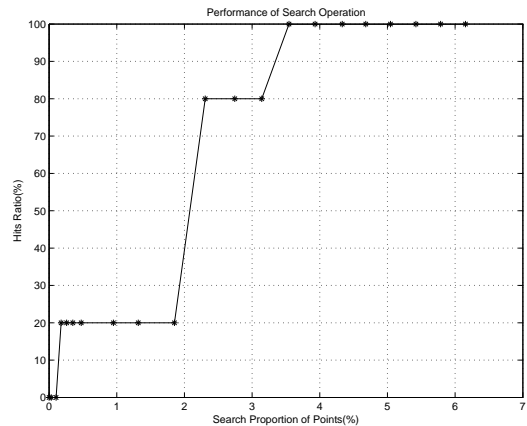}
  \caption{Search Proportions and Hits Ratio of 5NN on Synthetic Data}
  \label{fig:5NN synthetic search}
\end{figure}
Fig. \ref{fig:5NN synthetic search} shows the search operation on the synthetic data with 1 million tuples. Our algorithm performs well on this large sized synthetic data.

\subsection{Correctness under Team Context}
We perform the method under each team context on real dataset. The swap-out object identified with lower priority using our method is listed in Table \ref{tab:low priority}. The selected players are labelled according to their win-share of current play season.

\begin{table}
  \centering
  \caption{Lower Priority Swap-out Players}
    \begin{tabular}{ll}
    \toprule
    Team  & Low Priority Objects \\
    \midrule
    DEN   & 70,149,193,262,329,300 \\
    ORL   & 322,331 \\
    NYK   & 160,282,291,320,337 \\
    DAL   & 116,147 \\
    UTA   & 209,276,332 \\
    PHI   & 39,140,199,217,313 \\
    HOU   & 149,193,262,329,343 \\
    PHO   & 105,275,323 \\
    MIL   & 272,277,315 \\
    POR   & 266,350 \\
    \bottomrule
    \end{tabular}%
  \label{tab:low priority}%
\end{table}%

Take results of DEN listed in Table \ref{tab:den exchange} as an example. We labelled the players using their orders in the NBA League. Under this team context, we query only one nearest neighbour of each virtual object. Offset on \textit{x} dimension is the minimum value of current swap-out player and on \textit{y} dimension is its maximum value. In this experiment, we fixed our threshold as 0.02 as well.
\begin{table}[htbp]
  \centering
  \caption{Exchange Solution under DEN}
    \begin{tabular}{rrrc}
    \toprule
    Swap-out No.& Swap-in No.& Distance & results No. \\
    \midrule
    28    & 248   & 0.0235 & 159 \\
    52    & 279   & 0.0207 & 279 \\
    70    & 273   & 0.0534 & Low Priority \\
    97    & 135   & 0.0299 & 135 \\
    130   & 330   & 0.0308 & 330 \\
    142   & 185   & 0.031 & 185 \\
    149   & 217   & 0.0814 & Low Priority \\
    168   & 143   & 0.0386 & 143 \\
    193   & 217   & 0.1169 & Low Priority \\
    254   & 219   & 0.0835 & 219 \\
    262   & 217   & 0.0872 & Low Priority \\
    329   & 313   & 0.349 & Low Priority \\
    343   & 330   & 0.26  & Low Priority \\
    \bottomrule
    \end{tabular}%
  \label{tab:den exchange}%
\end{table}%

First three columns are results performed using exhaustive search. The last column is the results obtained using our algorithm. As described in Algorithm \ref{algo:search candidate}, the swap-out objects marked by us as "Low Priority" are not top ranked tuples under current team context. 

We can have a clear view that those objects with lower priority for searching is not top ranked ones comparing to others. Because although those objects are nearest neighbours of its virtual object, however, the distance in between is not small enough comparing to be ranked as an exchange solution. Only the exchange pairs with most minimum values are needed by us.

\subsection{A Comparison to Our Previous Work}
Our previous work proposed RTC* method in querying the nearest neighbours of current team context and then gave the final solution for an exchange plan.

\begin{figure}[H]
\center
     \includegraphics[scale=0.5]{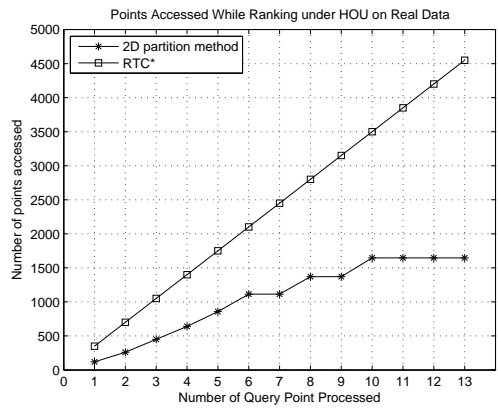}
     \caption{Comparision on Real Dataset}  \label{fig:compare us on real}
   \end{figure}

Fig.\ref{fig:compare us on real} shows the results of finding 5 top ranked tuples on real dataset under team context HOU. Since size of real dataset is small, although our 2D based partition method performs well, however, it still needs to query lots of data while handling the problem. Query points labelled 7, 10$\sim$13 are pre-pruned.

\begin{figure}[H]
\center
    \includegraphics[scale=0.5]{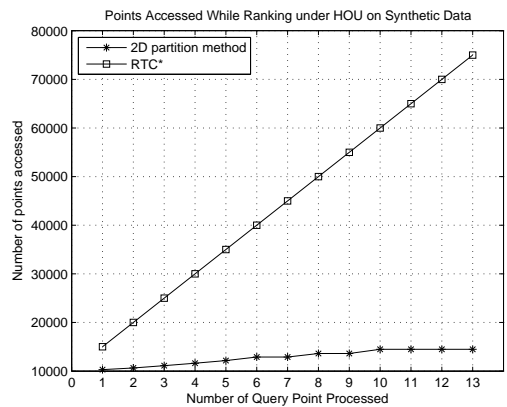}
 \caption{Comparision on Synthetic Dataset} \label{fig:compare our on synthe} 
\end{figure}

In Fig.\ref{fig:compare our on synthe}, we can observe that our algorithm in this paper outperforms RTC* in our previous work. So it can be generalize that method proposed in this paper can handle large dataset well.

\section{Conclusion and Future Work}\label{section:conclusion}
We propose a 2D partition based method in indexing the multidimensional data in this paper. We conduct experiments to show that our method performs NN query well on large sized skew data. Comparing to other existing techniques, this method can solve the problem of ranking under team context with effectiveness and efficiency.
In the future work, we plan to explore for reducing the number of window query aiming at reduce the overhead. Further, we plan to incorporate the cloud computing environment and consider to port the partition strategy.
\bibliographystyle{abbrvnat}
\bibliography{idea}
\end{document}